\RequirePackage{fix-cm}
\documentclass[smallcondensed]{svjour3}     % onecolumn (ditto)
\smartqed  % flush right qed marks, e.g. at end of proof
\usepackage{graphicx}
\usepackage{natbib}
\bibpunct[, ]{(}{)}{,}{a}{}{,}%
\def\bibhang{24pt}%
\def\makeheadbox{{%
\hbox to0pt{\vbox{\baselineskip=10dd\hrule\hbox
to\hsize{\vrule\kern3pt\vbox{\kern3pt
\hbox{\bfseries Scientometrics (2018) 117:1023-1040}
\hbox{This is a post-peer-review, pre-copyedit version of an article. The final}
\hbox{authenticated version is available online at: \href{https://doi.org/10.1007/s11192-018-2910-8}{doi.org/10.1007/s11192-018-2910-8}.}
\kern3pt}\hfil\kern3pt\vrule}\hrule}%
\hss}}}
\usepackage{dcolumn} % for Table
\usepackage{longtable}
\usepackage{ltxtable}
\usepackage{setspace}
\usepackage[ngerman, english]{babel}
\usepackage[latin1]{inputenc}
\usepackage[T1]{fontenc}
\usepackage{lmodern}
\usepackage{amssymb}
\usepackage{xcolor,graphicx,xspace,caption}
\usepackage{listings,eso-pic,subfigure}
\usepackage{lscape}
\usepackage{multirow}
\usepackage{enumerate}
\usepackage{amsmath}
\usepackage{pdfpages}
\usepackage{float}

\usepackage{dcolumn}
\newcolumntype{d}[1]{D{.}{.}{#1}}

\usepackage{fixltx2e}
\usepackage{hyperref}

\usepackage{multicol}

\begin{document}
	
\title{How to Measure the Performance of a Collaborative Research Center	\footnote{
	Financial support from the German Research Foundation (DFG) via Collaborative Research Center 649 ''Economic Risk'' and  International Research Training Group 1792 ''High Dimensional Nonstationary Time Series'', Humboldt-Universit\"{a}t zu Berlin, is gratefully acknowledged. We are thankful for the assistance provided by Nicole Hermann und Dominik Prugger.}
}
%\subtitle{}

\titlerunning{How to Measure the Performance of a CRC}        % if too long for running head

\author{Alona Zharova$^1$ %\item %\href{https://orcid.org/0000-0000-0000-0000}{\textcolor{orcidlogocol}{\aiOrcid} \hspace{2mm} orcid.org/0000-0000-0000-0000   
	\and
	\text{Janine Tellinger-Rice$^{1,2}$}  \and
	Wolfgang Karl H\"{a}rdle$^{1,3}$  
}

\authorrunning{Zharova, Tellinger-Rice and H\"{a}rdle} % if too long for running head

\institute{Alona Zharova \at
 \email{alona.zharova@hu-berlin.de}           %  \\
	\and
	{ }
	$^1$ School of Business and Economics, Humboldt-Universit\"{a}t zu Berlin, Germany\\
	$^2$ domino e.V., Germany\\ 
	$^3$ Singapore Management University, Singapore
}

\date{Received: 3 March 2018 / Accepted: 22 September 2018}

\maketitle
	
\begin{abstract}
	New Public Management helps universities and research institutions to perform in a highly competitive research environment. Evaluating publicly financed research improves transparency, helps in reflection and self-assessment, and provides information for strategic decision making. In this paper we provide empirical evidence using data from a Collaborative Research Center (CRC) on financial inputs and research output from 2005 to 2016. After selecting performance indicators suitable for a CRC, we describe  main properties of the data using visualization techniques. To study the relationship between the dimensions of research performance, we use a time fixed effects panel data model and fixed effects Poisson model. With the help of year dummy variables, we show how the pattern of research productivity changes over time  after controlling for staff and travel costs. The joint depiction of the time fixed effects and the research project's life cycle allows a better understanding of the development of the number of discussion papers over time.
			
	\vspace{0.2cm}
	\hspace{-0.5cm}\textbf{JEL} { C23 $\cdot$ C13 $\cdot$ M19 }
	%\PACS{PACS code1 \and PACS code2 \and more}
\end{abstract}

\newpage

\section{Introduction}
\label{sec1}

New Public Management (NPM) emerged in the 1980s (\citealt{Hood:91}) with the goal of improving efficiency and overall performance of public sector institutions by using business management approaches and models. NPM places a strong focus on permanent monitoring and evaluation of performance. Measuring research performance allows an analysis of the structural issues in science. It can thus facilitate the development of a scientific system and strengthen excellence in research.

This paper discusses Collaborative Research Centers (CRC) -- long-term university-based research institutions funded by the German Research Foundation (\citealt{DFG:18}). Evaluating publicly financed research results improves transparency, helps in reflection and self-assessment, and provides information for strategic decision making. Periodic monitoring of resource use and interim results allows CRC management to keep the finger on the pulse and to react to unfavourable phenomena promptly or to develop options for improvement; thereby, supporting success of the CRC.

There are numerous studies that concentrate on the evaluation of university research or research institutions in general (\citealt{Pastor:15}, \citealt{Berghe:98}). \cite{Lee:10} and \cite{Bolli:11} discuss performance measurements for departments and research units. \cite{Jansen:07} and \cite{Carayol:04} further investigate  performance indicators for research groups. However, a CRC differs from common research units or institutions, because of its interdisciplinary background. The performance indicators used for the evaluation of a CRC should be designed specifically for its needs and purposes in order to reflect the behaviour of involved research fields and other underlying characteristics.

In this paper we focus on a selection of performance indicators for intermediate and final evaluations suitable for broad applicability within CRCs and identifying a relationship between productivity and resource use of CRCs that may have implications for funding policy. 
The goals of this paper include: (i) selecting performance indicators suitable for a CRC; (ii) visualizing goals vs. results, societal impact and the interdisciplinarity structure of  research results of a CRC; (iii) analysis of a dependence structure between financial inputs and research output of a CRC and development of research productivity over time.

To achieve these objectives, we use twelve years (2005 -- 2016) of Collaborative Research Center 649 "Economic Risk" (CRC 649) data on 35 sub-projects. For each sub-project we observe yearly staff costs, travel costs and number of discussion papers (DPs). The life span of each sub-project varies, which results in an unbalanced panel.

\cite{Schroeder:14} indicate that the proposal for funding determines objectives for the research activity. To examine the correspondence between objectives and research results of the CRC, we carry out a semantic analysis of proposals and abstracts from published DPs. As a result, we find that both use 50\% of the same words.

Apart from research activity, a CRC has an impact on society through public events, transfer of knowledge or promotion of young researchers. For instance, young researchers usually perform  specific theoretical or practical research that is also used for their Ph.D. thesis. Collecting data on their further career helps to better understand this impact. With the help of a mosaic plot, we visualize three important dimensions of young researchers careers after receiving their Ph.D. within the CRC: gender, location and area of work. For example, we show that almost 70\% of young researchers who received their Ph.D. during CRC membership found later a job in academia.

Through a network analysis, we illustrate the interdisciplinarity structure of the research results and find out that most DPs were published in the fields of mathematical and quantitative methods, followed by financial economics, macroeconomics and monetary economics.

To study the relationship between research outcomes and funding for the CRC, we regress the number of DPs on staff and travel costs using sub-project-level data. With the help of year dummy variables added to the model, we show how the pattern of the sub-projects' productivity changed from 2005 to 2016 after controlling for staff and travel costs. Since the level of spending from the previous year and the preceding number of DPs may influence the current number of DPs, we additionally control for the lagged variables. The productivity of each sub-project may differ due to some heterogeneity or individual effects, such as the skills of a principal investigator (PI), average abilities or skills of researchers employed at the sub-project, or the specific behavior of a research field. For instance, working on a publication with one vs. more co-authors, writing in English vs. other languages, or publishing in books vs. articles may affect the research outcomes (\citealt{Zharova:17}). Therefore, we allow for the possibility of individual sub-project's effects. Considering the data structure, we apply a time fixed effects panel data (FE) model. Since the number of DPs is a count variable, we also apply a fixed effects Poisson (FEP) model.

We show that an increase of staff costs by 100\% leads to an expected increase in the number of DPs by roughly 43\% (FE) or 1.62 DPs (FEP). Travel costs have a diminishing effect on the number of DPs according to estimation results of the considered models. The previous level of both staff and travel costs negatively influence the number of DPs. We depict the estimates of coefficients of the dummy variables for years and find that the development trend corresponds with the stages of a project's life cycle. For instance, the most significant declines in the number of DPs take place during the stage of theoretical and empirical research, whereas the finalization stage corresponds with the growth in the number of published DPs.

The programmed \textsf{R} codes are available on the web-based repository hosting service and collaboration platform \href{https://github.com/AlonaZharova/CRC}{GitHub}.

The remainder of the paper is structured as follows. Literature review on performance indicators is presented in Section \ref{sec2}. Section \ref{sec3} describes the data and provides some preliminary descriptive analyses. Section \ref{sec4} introduces the methodology and shows empirical results. Finally, Section \ref{sec5} summarizes the results.

\section{Literature Review}
\label{sec2}

The combination of a peer-reviewed process and quantitative indicators is common practice in research performance assessment. The German Council of Science and Humanities (WR, germ. - Wissenschaftsrat) suggests evaluating the research institutions within three dimensions (research, promoting young researchers and knowledge transfer), which contain nine research performance criteria (\citealt{WR:04}). We select five criteria relevant to a CRC and provide a literature review on suitable indicators that may reflect the performance of the CRC.

1. \textit{Research quality} shows originality and novelty of research outputs, trustworthiness of methodology, impact and relevance for further research.

\begin{spacing}{}\footnotesize
	\begin{longtable}{p{2cm}|p{5.5cm}|p{3cm}}
		\hline\hline	
		Indicator & Definition & Literature \\ 
		\hline
		%			\textit{Quality of outputs} & &  \\
		%			Semantic analysis of goals and results & Comparison of goals and results & \\
		%			\hline
		\textit{Relative reception success} & & \\
		C\textsubscript{Pub} & Relation of total number of citations (NC\textsubscript{Pub}) to the total number of publications (N\textsubscript{Pub}) & \cite{WR:12}, \cite{Diem:13}, \cite{Donner:15} \\				
		%C\textsubscript{DP} & Relation of total number of citations (NC\textsubscript{DP}) to the total number of DP (N\textsubscript{Pub})  & \\	
		C\textsubscript{Pub}/FC\textsubscript{m} & Number of citations per publication in relation to the citation's average of the field & \cite{WR:12}, \cite{Abramo:11}, \cite{Moed:11}, \cite{Berghe:98} \\
		%C\textsubscript{DP}/FC\textsubscript{m} & Citations pro DP in relation to the citation's average of the field & \\
		C\textsubscript{Pub}/JC\textsubscript{m} & Number of citations per publication in relation to the citation's average of the journal & \cite{Moed:10}, \cite{WR:12} \\
		%C\textsubscript{DP}/JC\textsubscript{m} & Citations pro DP in relation to the citation's average of all DP & \\
		%N\textsubscript{DP->Pub} & Total number of publications that result from submitted DP & \\
		%Most cited publication per SP & & \cite{WR:12} \\
		\hline\hline
		\caption{Research quality.}
		\label{rq}
	\end{longtable}
\end{spacing}{}

2. \textit{Effectiveness} reflects the contribution of all sub-projects to the development of expertise in the research field within the CRC and beyond.

\begin{spacing}{}\footnotesize
	\begin{longtable}{p{2cm}|p{5.5cm}|p{3cm}}
		\hline\hline	
		Indicator & Definition & Literature \\ 
		\hline
		\textit{Research activity} & & \\
		N\textsubscript{Costs} & Total amount of the third party expenses (TPE) & \cite{WR:12}, \cite{Schmoch:09} \\
		N\textsubscript{Staff} & Total number of staff financed from third party funds (TPF) & \cite{Carayol:04}, \cite{WR:12} \\
		RA\textsubscript{unit} & Research activity of unit (sub-project, SP) -- multiplication of the total number of publications and the total number of citations of a unit with regard to the institutions-wide number of citations for the analyzed period (RA\textsubscript{SP}=N\textsubscript{Pub\textsubscript{SP}}*C\textsubscript{Pub\textsubscript{SP}}/C\textsubscript{Pub\textsubscript{CRC}}) & \cite{Pastor:15}\\
		\hline
		\textit{Research productivity} & & \\
		N\textsubscript{Pub} & Total number of publications & \cite{WR:12}, \cite{Abramo:11}, \cite{Diem:13}, \cite{Moed:11}, \cite{Hornbostel:91}  \\ 
		NC\textsubscript{Pub} & Total number of citations & \cite{WR:12}\\	
		%N\textsubscript{DP} & Total number of discussion papers & \\
		FN\textsubscript{Pub} & Fractional productivity -- total number of contributions to publications, where each contribution is a publication divided by the number of co-authors & \cite{Abramo:09}, \cite{Abramo:11}\\
		%FN\textsubscript{DP} & Fractional productivity -- total number of contributions to DP, where each contribution is a DP divided by the number of co-authors & \cite{Abramo:11}\\		
		ScS\textsubscript{Pub} & Scientific strength -- weighted sum of publications authored by each person, where the weights for each publication is the number of citations per publication in relation to the citation's average of the field (C\textsubscript{Pub}/FC\textsubscript{m}) & \cite{Abramo:11}, \cite{Abramo:09}\\
		%ScS\textsubscript{DP} & Scientific strength -- weighted sum of DP authored by each person, where the weights for each DP is the number of citations pro DP in relation to the citation's average of the field (DP\textsubscript{Pub}/FC\textsubscript{m}) & \\
		%Number if different journals & &  \\
		%N\textsubscript{Q} & Total number of Quantlets  & \\		
		\textit{h} & \textit{h}-index & \cite{Hirsch:05}, \cite{Bornmann:13}\\
		\hline
		\textit{Visibility of the CRC} & & \\
		AbsC\textsubscript{Pub} & Absolute citation count in the light of maximum citation count of a single publication (C\textsubscript{Pub\textsubscript{max}}) and the number of non-cited publications (N\textsubscript{ncPub}) & \cite{WR:12} \\
		%AbsC\textsubscript{DP} & Absolute citation count in the light of maximum citation count of a single publication (C\textsubscript{DP\textsubscript{max}}) and the number of non-cited publications (N\textsubscript{ncDP}) & \\
		\hline
		%	\textit{Interdisciplinarity} & & \\
		%	Networks & & \\
		%	\hline
		\textit{Reputation} & & \\
		& List of scientific prizes and awards &  \cite{Zheng:15}, \cite{WR:12}\\
		% CARI & Capacity of attracting resources index & \cite{Barra:16}\\
		%N\textsubscript{Guests} & Total number of guest researchers &  \cite{WR:12}\\
		\hline
		\textit{Professional activity} & & \cite{WR:12} \\
		& Editorships & \\
		& Review activities & \\
		& Editorial board memberships & \\
		& Academic functions & \\
		& Academic memberships &   \\
		& Organized conferences and workshops & \\
		\hline\hline
		\caption{Effectiveness.}
		\label{ie}
	\end{longtable}
\end{spacing}{}

3. The \textit{efficiency} criterion describes a quantity of research outputs in relation to a specific input, i.e. total costs, staff expenditures, number of staff, etc.

\begin{spacing}{}\footnotesize
	\begin{longtable}{p{2cm}|p{5.5cm}|p{3cm}}
		\hline\hline	
		Indicator & Definition & Literature \\ 
		\hline
		N\textsubscript{Pub}/N\textsubscript{Staff} & Relation of the number of publications (N\textsubscript{Pub}) to the number of research staff (N\textsubscript{Staff}) & \cite{Pastor:16}, \cite{WR:12}, \cite{Abramo:11}\\
		%N\textsubscript{DP}/N\textsubscript{Staff} & Relation of the total number of DP (N\textsubscript{DP}) to the total number of research staff (N\textsubscript{Staff}) & \\
		NC\textsubscript{Pub}/N\textsubscript{Staff} & Relation of the number of citations of publications (N\textsubscript{Pub}) to the number of research staff (N\textsubscript{Staff}) & \cite{WR:12}, \cite{Lee:10} \\
		%NC\textsubscript{DP}/N\textsubscript{Staff} & Relation of the total number of citations of DP (N\textsubscript{DP}) to the total number of research staff (N\textsubscript{Staff}) & \\
		N\textsubscript{Costs}/N\textsubscript{Staff} & Relation of the TPE to the total number of research staff (N\textsubscript{Staff})  &  \cite{WR:12}, \cite{Pastor:16}, \cite{Barra:16}\\
		%RP/CASI  & Research Productivity / Cost of academic staff index & \cite{Barra:16}\\
		\hline\hline
		\caption{Efficiency.}
		\label{e}
	\end{longtable}
\end{spacing}{}

4. \textit{Research enabling} relates to scientific activities that facilitate and support the research of young researchers.

\begin{spacing}{}\footnotesize
	\begin{longtable}{p{2cm}|p{5.5cm}|p{3cm}}
		\hline\hline	
		Indicator & Definition & Literature \\ 
		\hline
		\textit{Promotion of young } & & \\
		\textit{researchers} & & \\
		N\textsubscript{YR} & Total number of positions for young researchers &  \cite{WR:12}\\
		N\textsubscript{Ph.D.} & Total number of defended Ph.D. &  \cite{WR:12}, \cite{Diem:13}, \cite{Groezinger:06}, \cite{Schmoch:09} \\
		D\textsubscript{Ph.D.} & Average duration of Ph.D. study  &  \cite{WR:04}\\
		N\textsubscript{Pub\textsubscript{Ph.D.}} & Total number of publications by young researchers & \cite{WR:04}\\
		%		Completed habilitations & &  \cite{WR:12}\\
		%		Completed habilitations by females & &  \cite{WR:12}\\
		& List of awards and prizes of young researchers &  \cite{WR:12}\\
		& List of calls and appointments for young researchers &  \cite{WR:12}\\
		\hline
		%	\textit{Infrastructures and networks} & & \\ Infrastructures & [Data Center Services, Databases] & \cite{WR:12} \\
		%	Networks and associations & & \cite{WR:12} \\
		%	& List of organized conferences and workshops &  \cite{WR:12}\\
		\hline\hline
		\caption{Research Enabling / Promotion of young researchers.}
		\label{re}
	\end{longtable}
\end{spacing}{}

5. \textit{Knowledge transfer} defines the transfer of research results and products or  distribution of knowledge.

\begin{spacing}{}\footnotesize
	\begin{longtable}{p{2cm}|p{5.5cm}|p{3cm}}
		\hline\hline	
		Indicator & Definition & Literature \\ 
		\hline
		N\textsubscript{Pat} & Number of patents & \cite{WR:11}, \cite{Carayol:04} \\
		& List of Transfer projects & \\
		& List of activities in public relations & \cite{WR:12} \\
		& List of research products and teaching materials & \cite{WR:12} \\
		\hline\hline
		\caption{Knowledge Transfer.}
		\label{kt}
	\end{longtable}
\end{spacing}{}

\section{Data}
\label{sec3}

Collaborative Research Centers (CRC) are interdisciplinary research institutions financed through the German Research Foundation (germ. - Deutsche Forschungsgemeinschaft, DFG). The goal of a CRC is to pursue interdisciplinary innovative research by bringing together scholars from different research fields within multiple research projects, also called sub-projects. The classical CRC consolidates cooperation between several universities or non-university research institutions with at least 60\% of all sub-projects based in the coordinating university (\citealt{DFG:18}).

CRCs are granted for four years and depending on the results of the interim evaluations can be prolonged twice for a maximum period of twelve years. 
During the assessment each sub-project undergoes a critical appraisal. Depending on a change in research program or staff turnover (professors), a CRC can also submit proposals for new sub-projects. As a result, the number of research projects may vary between phases.

In this paper we provide empirical evidence using data from a Collaborative Research Center 649 "Economic Risk" (hereinafter referred to as the CRC). The CRC was launched in 2005 for a four-year term and extended twice, for a total life span of twelve years. As an interdisciplinary research center, it combined economics, mathematics and statistics and pursued research within three primary areas: i) microeconomics, in particular individual and contractual answers to risk; ii) quantitative projects, in particular financial markets and risk assessment; iii) macroeconomic risks. For more information, we refer to the website of the CRC (\citealt{CRC:16}).

\begin{figure}[htp]									
	\begin{center}
		\includegraphics[scale = 0.57]{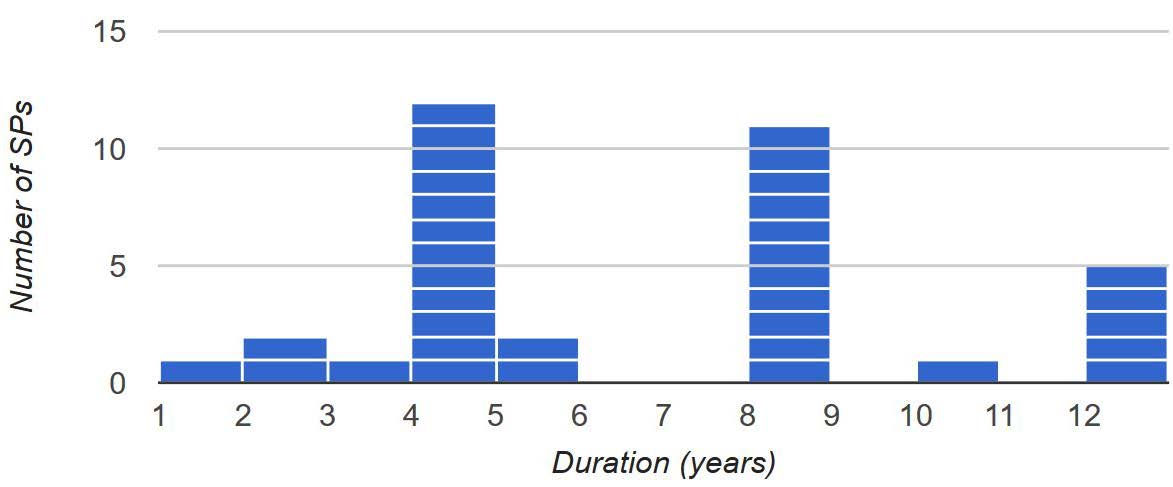}
		\caption{Distribution of sub-projects (SP) over life span in years.
		}
		\label{CRChissp}
	\end{center}
\end{figure}

The total number of the CRC sub-projects within three four-year phases is 35, but the number of sub-projects per phase varies from 16 to 21. Since the sub-projects of the CRC have different life periods, the data set does not have the observations for all years that indicates an unbalanced panel, see Figure \ref{CRChissp}. The main reason for the panel being unbalanced is the attrition of sub-projects, as a result of research project's termination or the leave of principal investigators to other universities, and the establishment of new research projects during the prolongation phases. For instance, twelve sub-projects had a life cycle of four years, eleven sub-projects lasted for eight years and five sub-projects  existed twelve years (see Figure \ref{CRChissp}).

Principal investigators (PIs) lead sub-projects. From 35 sub-projects 83\% have one PI and 17\% have two PIs. Since three PIs participate in two sub-projects, the CRC counts 38 PIs in total over twelve years. PIs of all three academic ranks participate in the CRC: full professors (76\%), junior professors (19\%) and postdoctoral researchers (5\%).

The CRC uses 62\% of resources on average to finance the research staff working within sub-projects, in particular doctoral (Docs) and postdoctoral (PostDocs) researchers. In addition, all members of the CRC may use its central funds for travel costs, organizing conferences and workshops, inviting guest lecturers and researchers, gender equality etc.

\begin{figure}[htp]									
	\begin{center}
		\includegraphics[scale = 0.46]{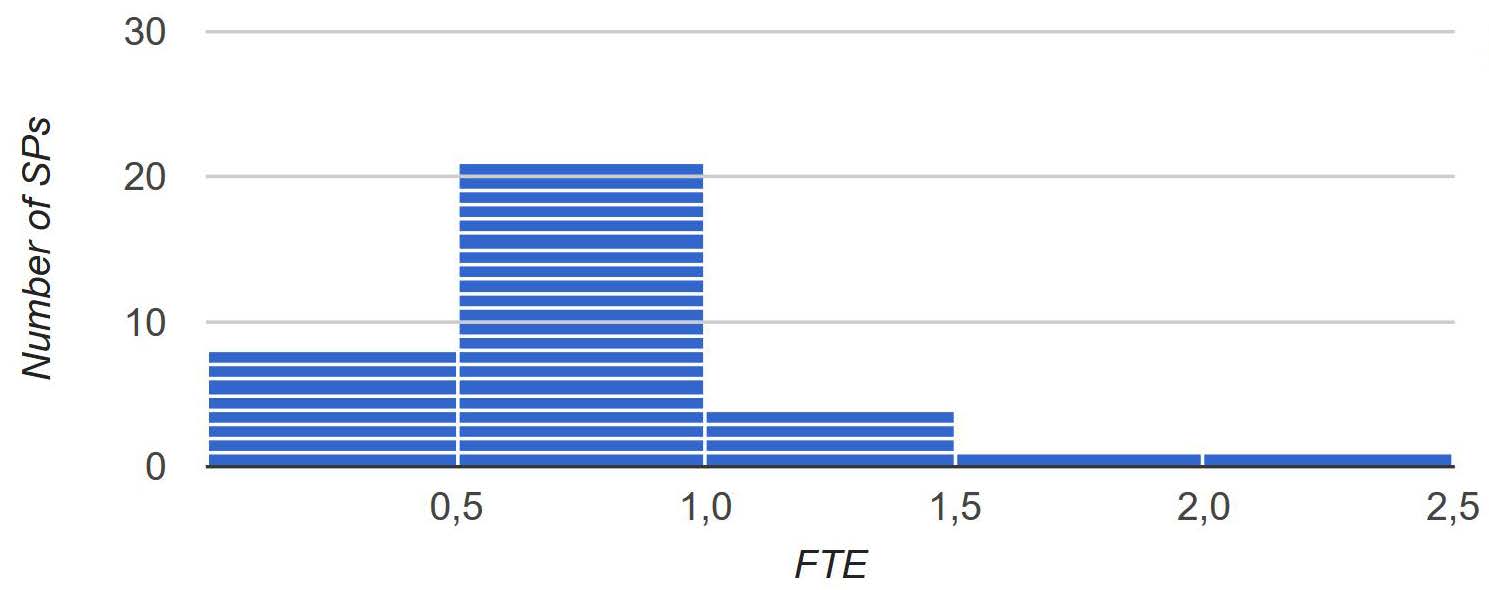}
		\caption{Distribution of sub-projects according to the number of research staff (in FTE per year).
		}
		\label{CRChissp_fte}
	\end{center}
\end{figure}

The amount of research staff working within sub-projects differs, depending on the scope and complexity of the research program. Each sub-project counts from 0.5 to 2.5 full-time equivalents (FTEs) of researcher positions per year. The FTEs are often split and used to hire more research staff, i.e. 2 researchers with 50\% financing, or to top up researchers that are already employed and who are financed by other sources. Figure \ref{CRChissp_fte} shows the distribution of sub-projects according to the number of FTEs per year. For instance, 21 sub-projects have one FTE per year on average, eight sub-projects hire staff on 0.5 FTEs, four sub-projects use 1.5 FTEs and two sub-projects have each 2 and 2.5 FTEs.

\begin{figure}[htp]										
	\begin{center}
		\includegraphics[scale = 0.6]{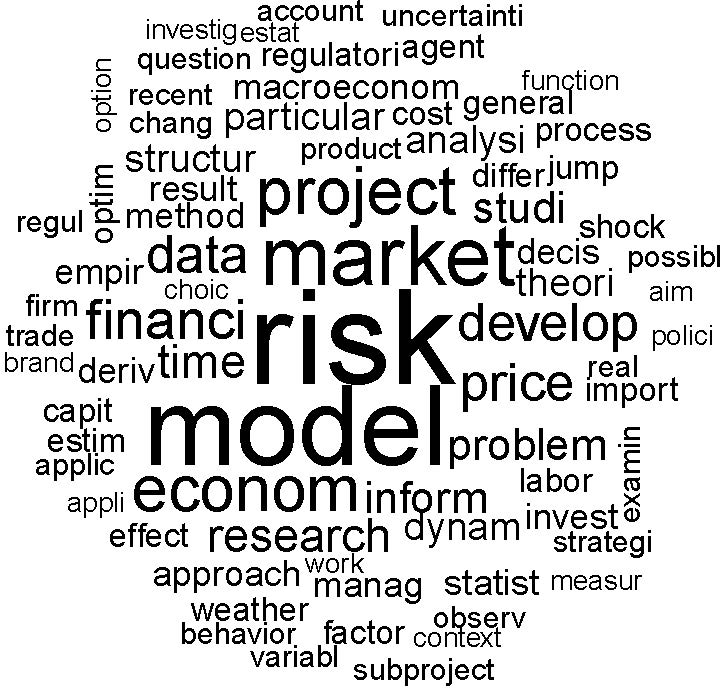}
		\hspace{0.5cm}
		\includegraphics[scale = 0.6]{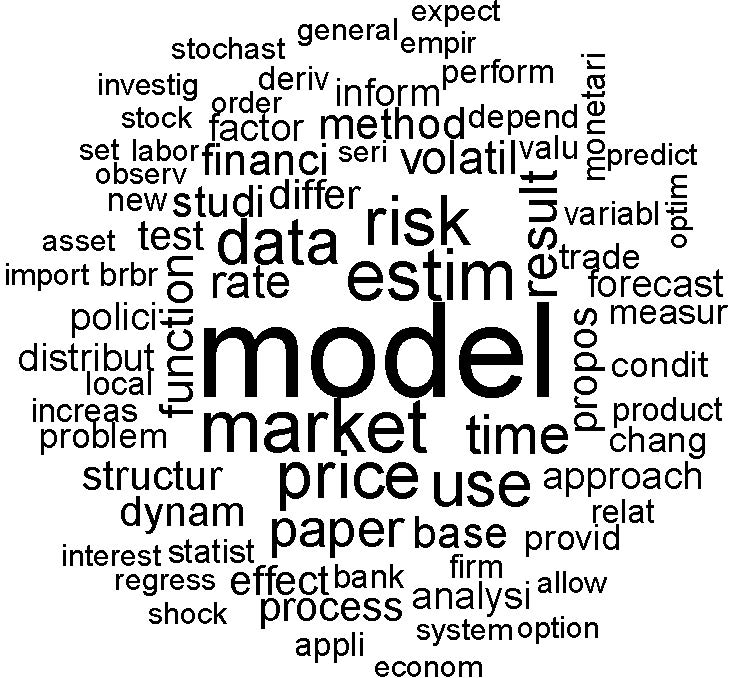}
		\caption{Semantic analysis of goals (left; 61 summaries from sub-projects of three proposals for the CRC) vs. results (right; 771 abstracts from DP).
		}
		\label{Semant}
	\end{center}
\end{figure}

In this paper we use data from annual financial reports, internal publications' and discussion papers' (DPs) databases and CRC's newsletter. Additional insight is gathered from the texts of one proposal for a launch and two proposals for a prolongation of the CRC 649 (2005--2008, 2009--2012, 2013--2016) which were submitted to the DFG. On the one hand, one can see such proposals as goals that the CRC sets for each period. On the other hand, the published DPs encompass the achieved results of the research activity. We undertake a semantic analysis on both informational sources, i.e. 61 summaries of sub-projects from three proposals and abstracts of 771 DPs. The two word clouds of the top 75 keywords are illustrated in Figure \ref{Semant}. We find that both use 50\% of the same words. The different size of the same words, for instance the word "risk", indicates that the number of times the word is mentioned in the proposals and abstracts differs.

\begin{figure}[H]
	\begin{center}
		\includegraphics[scale = 0.32]{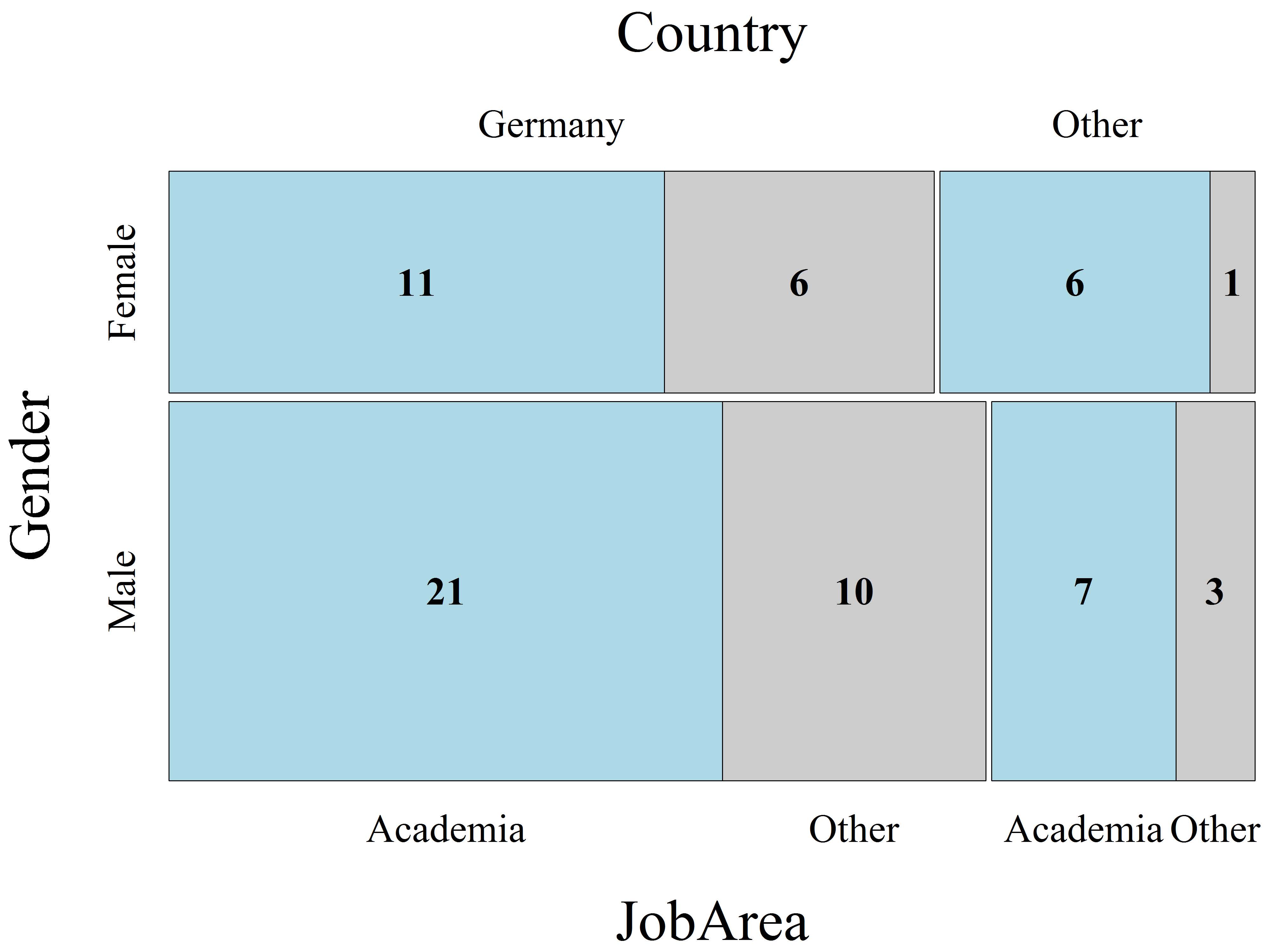}
		\caption{Mosaic plot of job type, location and gender of 65 CRC members who received their Ph.D. between 2005 and 2016 (as of Dec 2016)
		}
		\label{Mosaic}
	\end{center}
\end{figure}

One of the primary goals of a CRC is the high-quality instruction, supervision and support of young researchers. The common result of this process is a Ph.D. defence. Collecting data on the further career of the young researchers helps to better understand the impact on society. For instance, one may wonder how many females that worked and defended their Ph.D. thesis at the CRC are afterward working in academia in Germany? To visualize such data we use a mosaic plot in Figure \ref{Mosaic}.

The vertical axis splits the individuals according to their gender. The data are further divided into two groups on the upper horizontal axis according to the location of the job. The lower horizontal axis shows how many people received a contract in academia or other fields.
The width and height of each segment represent the number of observations within each group.
Consider the 65 members of the CRC that received their Ph.D. from 2005 to 2016. There are 11 female researchers that received jobs in academia in Germany and 6 in other countries. For males that stayed in academia, the number is 21 for Germany and 7 for other countries. This means that almost 70\% of young researchers who received their Ph.D. during CRC membership found later a job in an academic institution.

\begin{figure}[H]
	\begin{center}
		\includegraphics[scale = 0.5]{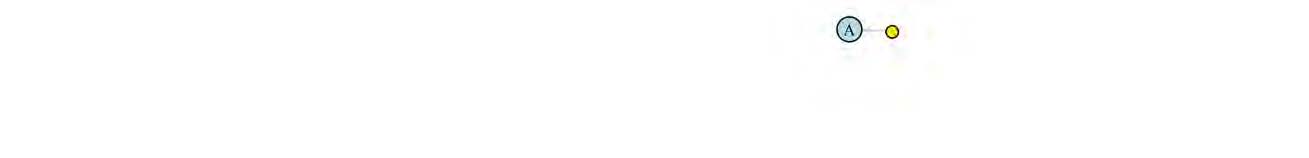}
		\includegraphics[scale = 0.5]{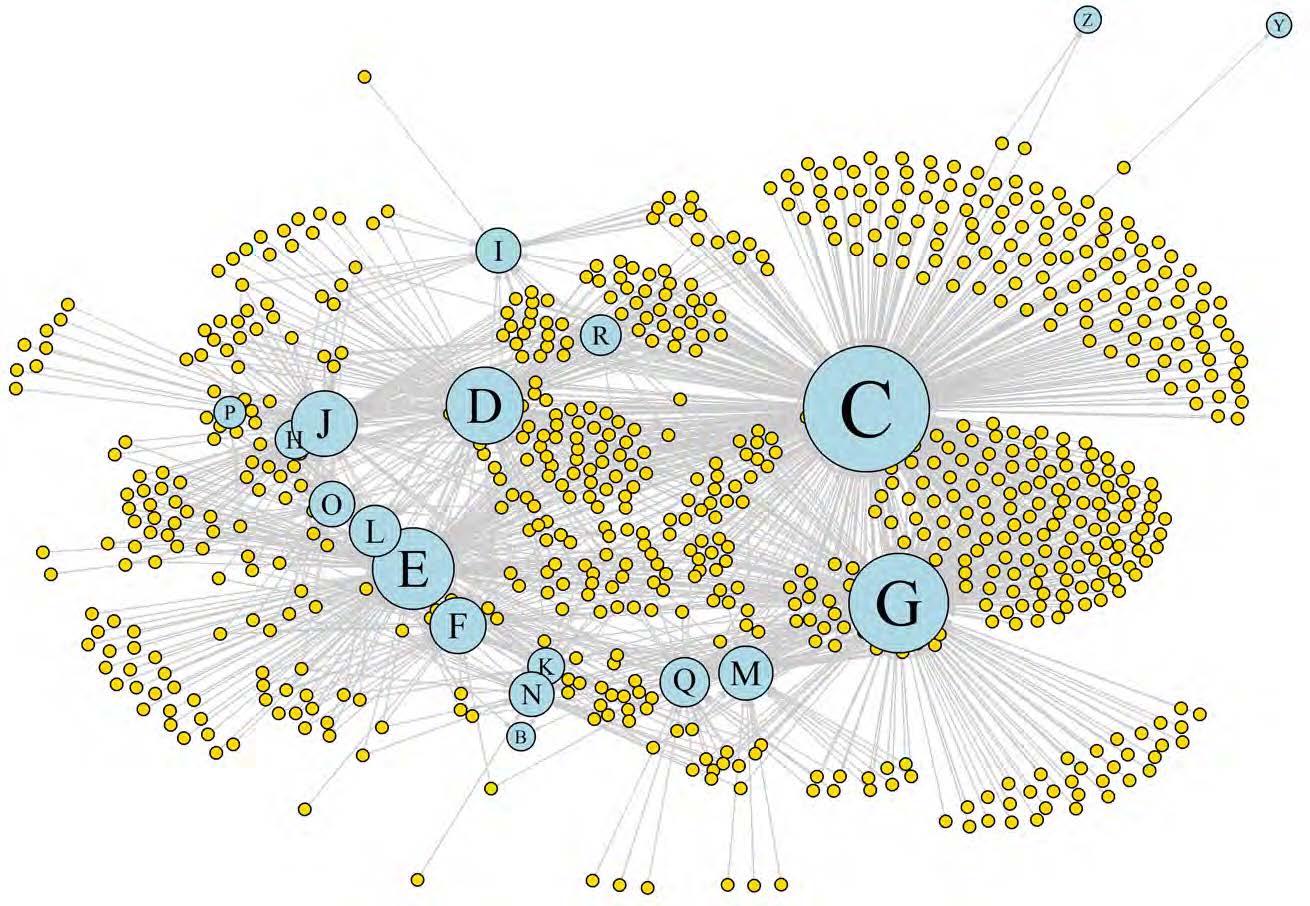}
		\caption{Network of 760 discussion papers (yellow) and 20 JEL codes (blue) published from 2005 to 2016.		
		}
		\label{CRCnetjel}
	\end{center}
\end{figure}

The proportion of 36.9\% of female researchers is quite low in comparison to 50.4\% for female doctoral students within CRCs in social sciences and humanities, but higher than 25.7\% within CRC in mathematical and natural sciences (\citealt{DFG:17}). However, since the CRC pursued interdisciplinary research in both social and mathematical sciences, the CRC proportion corresponds to the value in-between. As a part of the communication processes with alumni and mentoring of CRC young researchers, the CRC invited its former members who got promoted in academia as guest lecturers for CRC seminars or as guest researchers to work on papers jointly with PIs and/or younger CRC generations. 

In order to understand if the intended interdisciplinarity occurred, we analyze DPs that serve as an outcome of the CRC research activity. Almost each DP has codes indicating subject fields according to the Journal of Economic Literature (JEL) classification in the economic sciences (see \citealt{JEL:18}).  

We show the network of collaborating disciplines in Figure \ref{CRCnetjel}. The small gold circles introduce the DPs, whereas the nodes leading to the bigger blue circles indicate the JEL code of the corresponding research area. The size of each blue circle reflects the relative number of references to DPs. The explanation of JEL codes is given in Table \ref{JELcodes}. For instance, most of the DPs were published in the C area, i.e. mathematical and quantitative methods. They are followed by G (financial economics),  E (macroeconomics and monetary economics) and D (microeconomics). These four fields with higher research output correspond to the three primary areas of the CRC. Note that the DPs that involve research in more than one field are connected to two or more JEL codes simultaneously. This confirms the interdisciplinary character of the CRC research output.

\begin{longtable}[H]{r|p{10cm}}
	\hline\hline
	{Code} & {Research field}\\
	\hline 
	A & General Economics and Teaching \\
	B & History of Economic Thought,  Methodology, and Heterodox Approaches \\
	C & Mathematical and Quantitative Methods \\
	D & Microeconomics \\
	E & Macroeconomics and Monetary Economics \\
	F & International Economics \\
	G & Financial Economics \\
	H & Public Economics \\
	I & Health, Education, and Welfare \\
	J & Labor and Demographic Economics \\
	K & Law and Economics \\
	L & Industrial Organization \\
	M & Business Administration and Business Economics / Marketing / Accounting / Personnel Economics \\
	N & Economic History \\
	O & Economic Development, Innovation,  Technological Change, and Growth \\
	P & Economic Systems \\
	Q & Agricultural and Natural Resource  Economics / Environmental and Ecological Economics \\
	R & Urban, Rural, Regional, Real Estate,  and Transportation Economics \\
	Y & Miscellaneous Categories \\
	Z & Other Special Topics \\
	\hline \hline
	\caption{JEL Classification System}
	\label{JELcodes}		
\end{longtable}

\begin{figure}[htp]									
	\begin{center}
		\includegraphics[scale = 0.415]{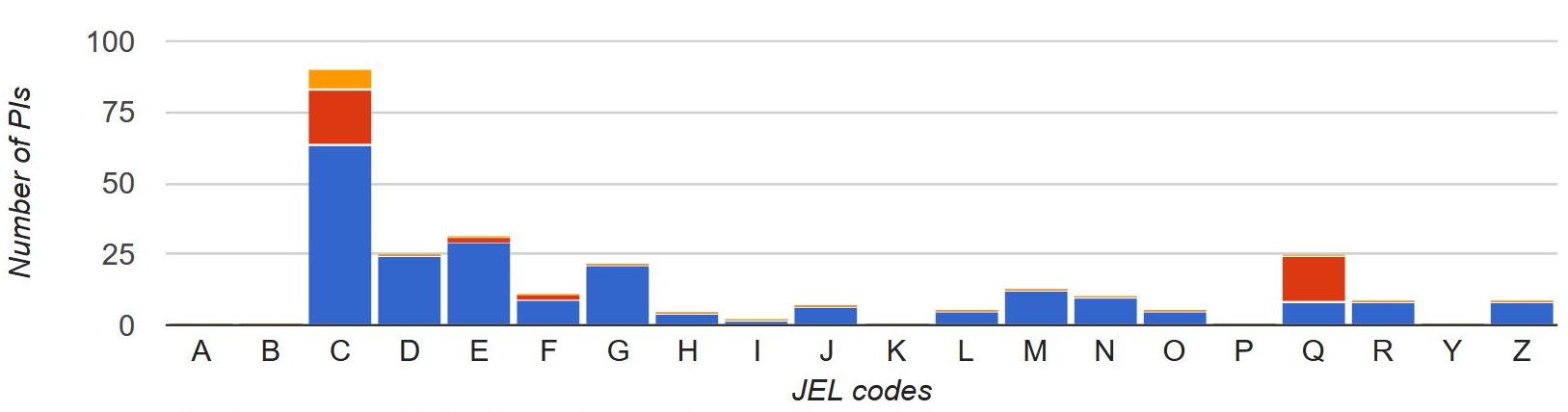}
		\caption{Cumulative number of PIs (in PI years; full professors - blue, junior professors - red, postdoctoral researchers - orange) from 2005 to 2016 (weighted by the number of research fields and sub-projects) with expertise in corresponding JEL research fields.
		}
		\label{CRCstaffFieldsPI}
	\end{center}
\end{figure}

\begin{figure}[htp]									
	\begin{center}
		\includegraphics[scale = 0.415]{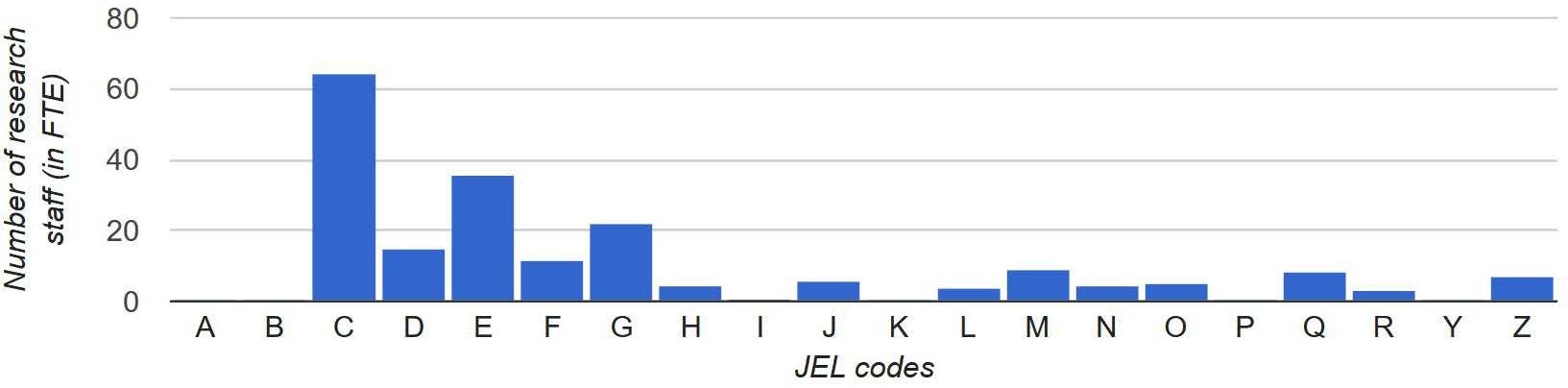}
		\caption{Cumulative number of research staff in FTE (in staff years; weighted by the number of research fields) from 2005 to 2016 working within corresponding JEL research areas.
		}
		\label{CRCstaffFieldsRstaff}
	\end{center}
\end{figure}

One more factor influencing the variability of the number of DPs across research fields is the area of expertise of PIs and research staff. Figure \ref{CRCstaffFieldsPI} shows the cumulative number of PIs within their areas of expertise and Figure \ref{CRCstaffFieldsRstaff} depicts the cumulative number of CRC research staff (in FTE) working within same research areas for twelve years. Since the attrition of some sub-projects and establishment of new ones influences the availability of PIs and research staff and accordingly their expertise within the CRC life cycle, we use cumulative numbers. We also use weights for the number of the sub-projects and expertise areas for each PI to equalize the total time available for research. For example, the PI who is an expert in four research areas receives 0.25 for each JEL code and the PI who leads two sub-projects has 0.5 for the distribution within JEL areas of each project. 

Figures \ref{CRCstaffFieldsPI} and \ref{CRCstaffFieldsRstaff} show, for instance, that the area D reveals 24 years of PIs expertise and 15 years of research staff (in FTE) work. Both figures provide evidence that the most expertise is concentrated within the area C, followed by E, D, G and Q. This also explains the concentration of research output within corresponding JEL areas in Figure \ref{CRCnetjel}. The correlation between the number of DPs and number of PIs specializing in the same JEL areas is 93.8\%  (95\% for full professors only), whereas the correlation between the number of DPs and the amount of research staff (in FTE) working within same fields is 95.1\%.

\section{Analysis of Research Productivity}
\label{sec4}

The observed time series across the same sub-projects indicate the longitudinal or panel structure of the data. To investigate the relationship between the input and the output variables, we use the methods designed for panels.

\subsection{Methodology}

The basic framework for the panel data analysis shows the  %unobserved effects 
model (\citealt{Wooldridge:02}):

\begin{equation}
y_{i} = \beta X_{i} + u_{i},  
\hspace{0.5cm} i=1,\ldots, K,
\label{PD}
\end{equation}
where $y_{i} = (y_{i1}, \ldots,y_{iT})^{\top}$  is a ($1\times T$) vector of observations for $t=1,2, \ldots,T$, $X_{i} = (x_{i1}^{\top}, \ldots, x_{iT}^{\top})^{\top}$ is a ($K\times T$) matrix of observations, $\beta$ is a  ($K\times 1$) vector of coefficients and $u_{i}$ is a ($1\times T$) vector of unobservables. 

The unobserved sub-project's effect may contain such factors as publishing behavior in a research field, average researchers' abilities or skills of principal investigators of sub-projects that should be roughly constant over time. 

We allow for arbitrary correlation between the unobserved sub-project's heterogeneity or fixed effects $c_i$ and the observed explanatory variables $x_{it}$ and, therefore, use the fixed effects model for each $i$ (\citealt{Wooldridge:16}):

\begin{equation}
y_{it} = \beta_1 x_{it1} + \ldots + \beta_k x_{itk} + c_{i} + u_{it}, \hspace{0.5cm}  t=1,2,\ldots,T, 
\hspace{0.5cm} i=1,2,\ldots,K,
\label{FE}
\end{equation}
% the parameters to estimate
%\begin{equation}
%y_{t} = \alpha + \sum_{j=1}^{p}A_jy_{t-j} + \varepsilon_{t},  ~\varepsilon_{t} \sim N(0,\Sigma),~(\varepsilon_t) ~\text{i.i.d.}
%\label{VAR}
%\end{equation}
where $y_{it}$ includes dependent variables and $x_{it}$ independent variables for individual $i$ at time $t$, $\beta_1, \ldots, \beta_k$ are the unknown coefficients, $c_i$ is individual effect or individual heterogeneity and $u_{it}$ are idiosyncratic errors that change across individuals $i$ and time $t$. 

The fixed effects estimator (or the within estimator) is obtained as the pooled OLS estimator on the time-demeaned variables. The strict exogeneity assumption on explanatory variables, $\operatorname{E}(u_{it} |  \operatorname{\textbf{X}}_i, c_i) = 0$, provides that the fixed effects estimator is unbiased (\citealp{Wooldridge:16}). As the number of sub-projects (clusters) is large, statistical inference after OLS should be based on cluster-robust standard errors to account for heteroscedasticity and within-panel serial correlation (\citealt{Cameron:15}).

Next, we are interested in the pattern of sub-projects' productivity, i.e. number of produced discussion papers, in different time periods. For this purpose we use time fixed effects that change over time but are constant across sub-projects. We include the dummy variables for $T-1$ years to avoid the multicollinearity. Usually  the first year is selected as a base year.  The time fixed effects model (FE) is (\citealt{Stock:03}):

\begin{equation}
y_{it} = \beta_1 x_{it1} + \ldots + \beta_k x_{itk} +
\delta_1 + \delta_2 D_2 + \ldots + \delta_T D_T
+ c_{i} + u_{it}, 
\label{FEt}
\end{equation}
where $D_2, \ldots, D_T$ are time effects and $\delta_1, \ldots, \delta_T$ are the parameters to estimate.

%For more detailed information, please refer to W16 and SW03.

When the dependent variable involves count data, it has a Poisson distribution instead of a normal distribution. \cite{Hausman:84} introduce a fixed effects Poisson model (FEP) as:

%assumes
%\begin{equation}
%y_{it} | x_i, a_i \sim Poisson(a_i \mu (x_{it}, \beta_0)), \hspace{0.5cm}  t=1,2,\ldots,T, 
%\label{}
%\end{equation}

\begin{equation}
\operatorname{E}( y_{it} | x_{i}, a_i) = a_i \mu (x_{it}, \beta_0), \hspace{0.5cm}  t=1,2,\ldots,T, 
\label{}
\end{equation}
where $\beta_0$ is a ($1 \times K$) vector of unknown parameters to be estimated and $\mu$ is the conditional mean. \cite{Wooldridge:99} further derives a consistent estimator for FEP using a quasi-conditional maximum likelihood estimator (QCMLE).

\subsection{Empirical Results}

%\begin{equation}
%y_{it\_{ndp}} = \beta_{stac} x_{it\_{stac}} + \beta_{trac} x_{it\_ {trac}} + \delta_{2005} + \delta_{2006} D_{2006} + \ldots + \delta_{2016} D_{2016} + c_{i} + u_{it}, 
%\label{FEse}
%\end{equation}
%\begin{equation*}
%\hspace{0.3cm}  t=2005, 2006, \ldots, 2016, 
%\hspace{0.3cm} i=1,2,\ldots,32,
%\end{equation*}

Before presenting the estimates, we explain some specifications of the model. 
Since the yearly staff and travel costs are in nominal Euros, a slight increase may happen due to inflation.
One possibility to deal with this is an adjustment using a Consumer Price Index (CPI). Another way to track the effect of real spendings is the use of a logarithmic form. The interpretation of the estimation results is then done using the level-log model. Here we use the second approach.

Table \ref{EstRes_nDP} presents the 
results of FE (1) and (2), and FEP (3) and (4) models for the number of DP as a dependent variable. 
The parameters of interest are staff costs $\beta_{logStaffCosts}$, travel costs $\beta_{logTravelCosts}$ and year-specific influence $\delta_{year}$. % and the unobserved SP effect $const$
We also include lagged variables into the models (2) and (4), since the current number of research outputs may be affected by the previous number of publication and invested funds in economic sciences and mathematics (\citealt{Zharova:17}). The models (2) and (4) encompass the number of DPs $\beta_{nDP_{t-1}}$, staff costs $\beta_{logStaffCosts}$ and travel costs $\beta_{logTravelCosts}$ in the time $t-1$. The intercept $const$ is the average of individual effects $c_i$ across all sub-projects that is reported by Stata.
We use cluster-robust standard errors to account for heteroscedasticity. The significance level of all estimates decreases as a result of standard error adjustment (\citealt{Wooldridge:16}).

\begin{table}[H]
	\centering
	%\scriptsize{
	\fontsize{9}{10}\selectfont
	%\begin{tabular}{p{3.5cm}|p{1.5cm}p{1.5cm}|p{1.5cm}p{1.5cm}}
	\begin{tabular}{p{3.5cm}|d{3.2}d{3.2}|d{3.2}d{3.2}}
		%\toprule
		\hline
		\hline
		Dependent variable: $nDP$&\multicolumn{2}{c|}{FE model}&\multicolumn{2}{c}{FEP model}\\
		&\multicolumn{1}{c}{(1)}&\multicolumn{1}{c|}{(2)}&\multicolumn{1}{c}{(3)}&\multicolumn{1}{c}{(4)}\\
		\hline
		$\beta_{logStaffCosts}$  &    1.38  ^{**}    &    1.62  ^{*}    &    0.47  ^{***}    &    0.43  ^{**}    \\
		&  (  0.61    )  &  (  0.88    )  &  (  0.12    )  &  (  0.19    )  \\
		$\beta_{logTravelCosts}$  &    -0.94  ^{*}    &    -0.34      &    -0.22  ^{**}    &    -0.04      \\
		&  (  0.55    )  &  (  0.47    )  &  (  0.10    )  &  (  0.09    )  \\
		$\delta_{2006}$  &    1.61      &    1.92      &    0.25      &    0      \\
		&  (  1.36    )  &  (  1.61    )  &  (  0.26    )  &    \multicolumn{1}{c}{(\textit{omit.})}      \\
		$\delta_{2007}$  &    -1.20      &    -2.55      &    -0.30      &    -0.98  ^{***}    \\
		&  (  1.38    )  &  (  2.46    )  &  (  0.31    )  &  (  0.25    )  \\
		$\delta_{2008}$  &    -0.95      &    -2.03      &    -0.23      &    -0.97  ^{***}    \\
		&  (  1.30    )  &  (  2.10    )  &  (  0.32    )  &  (  0.36    )  \\
		$\delta_{2009}$  &    -2.05  ^{*}    &    -3.16      &    -0.54  ^{*}    &    -1.20  ^{***}    \\
		&  (  1.13    )  &  (  1.98    )  &  (  0.33    )  &  (  0.23    )  \\
		$\delta_{2010}$  &    -1.93  ^{*}    &    -2.13      &    -0.51  ^{*}    &    -1.03  ^{***}    \\
		&  (  1.14    )  &  (  2.68    )  &  (  0.30    )  &  (  0.31    )  \\
		$\delta_{2011}$  &    1.10      &    0      &    0.33  ^{*}    &    0      \\
		&  (  0.70    )  &    \multicolumn{1}{c|}{(\textit{omit.})}      &  (  0.20    )  &    \multicolumn{1}{c}{(\textit{omit.})}     \\
		$\delta_{2012}$  &    -2.79  ^{*}    &    -3.60  ^{*}    &    -0.71  ^{**}    &    -1.90  ^{***}    \\
		&  (  1.46    )  &  (  1.78    )  &  (  0.34    )  &  (  0.20    )  \\
		$\delta_{2013}$  &    -2.98  ^{**}    &    -3.18      &    -0.80  ^{**}    &    -1.32  ^{***}    \\
		&  (  1.30    )  &  (  2.52    )  &  (  0.32    )  &  (  0.41    )  \\
		$\delta_{2014}$  &    -1.36      &    -1.73      &    -0.44      &    -0.99  ^{***}    \\
		&  (  0.95    )  &  (  1.61    )  &  (  0.27    )  &  (  0.37    )  \\
		$\delta_{2015}$  &    -2.55  ^{**}    &    -1.90      &    -0.74  ^{**}    &    -1.02  ^{***}    \\
		&  (  1.17    )  &  (  1.77    )  &  (  0.33    )  &  (  0.31    )  \\
		$\delta_{2016}$  &    -0.30      &    0      &    -0.31      &    -0.69  ^{*}    \\
		&  (  1.79    )  &  \multicolumn{1}{c|}{(\textit{omit.})}  &  (  0.36    )  &  (  0.41    )  \\
		$const$  &    -2.37      &    0.05      &          &         \\
		&  (  5.29    )  &  (  10.09    )  &          &          \\
		$\beta_{nDP_{t-1}}$  &          &    0.02      &          &    -0.01  ^{*}    \\
		&          &  (  0.16    )  &          &  (  0.03    )  \\
		$\beta_{logStaffCosts_{t-1}}$  &          &    -0.66      &          &    -0.25      \\
		&          &  (  0.59    )  &          &  (  0.23    )  \\
		$\beta_{logTravelCosts_{t-1}}$  &          &    -0.21      &          &    -0.02      \\
		&          &  (  0.58    )  &          &  (  0.13    )  \\
		\hline
		$R^2$ &  0.20   &  0.21   &     &     \\
		AIC &  706   &  437   &  463   &  253   \\
		BIC &  742   &  469   &  501   &  258   \\
		\hline 
		\hline
		%\bottomrule
	\end{tabular}
	%	}
	\caption{Estimation results for time fixed effects (within) regression (models (1) and (2)) and fixed effects Poisson regression (models (3) and (4)) with number of DP ($nDP$) as the dependent variable and with robust standard errors adjusted for clusters in sub-projects. ***, ** and * indicate a statistical significance at 1\%, 5\% and 10\% level, respectively. Standard deviation is provided in brackets. } 
	\label{EstRes_nDP}
\end{table}

In (2) and (4) two years were omitted because of collinearity. In (3) five observations were dropped out of the analysis because there was only one observation per group. Performing analysis on unbalanced data slightly increases the estimated effects of considered variables, but the general idea remains unchanged (\citealt{Wooldridge:16}).

In the model (1) we see the positive, significant effect of staff costs on the number of DPs. 1.38/100 is the unit change in $n{DP}$ when staff expenses increase by 1\%. In other words, a 100\% increase in staff costs leads to an increase in the number of DPs by 1.38. Similarly, the model (2) shows that a 100\% increase in staff costs increases the number of DPs by 1.62, holding other variables constant. The fit of the FE models in (1) and (2) in Table \ref{EstRes_nDP} with $nDP$ as the dependent variable is almost the same, indicating that including lagged variables does not significantly improve the model.

The FEP estimates have a different interpretation. For instance, the coefficient on $\beta_{logStaffCosts}$ shows that a rise of staff costs by 100\% leads to an increase of the number of DPs by 47\% and 43\% for models (3) and (4) correspondingly. The coefficients on staff costs estimates for four models in Table \ref{EstRes_nDP} are significant at 1\% to 10\% level. The influence of previous values of staff costs on the number of DPs is negative and insignificant.

Travel costs have a diminishing effect on the number of DPs according to estimation results of considered models. The coefficient on $\beta_{logTravelCosts}$ implies that, if we increase the travel costs by 100\%, we expect the number of DP to decrease by 0.94 DP due to FE model (1). The Poisson coefficient in (3) means that an increase in $logTravelCosts$ by 10\% decreases $nDP$ by 2\% (0.22$\times$0.10). 

The coefficients on the year dummy variables reveal how the average productivity of sub-projects changes over time. As 2005 is selected as the base year, it is not reported with a coefficient. The coefficient on $\delta_{2006}$ in model (1) shows that, on average, 1.6 DPs are attributed to the year effect of 2006 holding other factors fixed. In Poisson case (3) one suggests that the expected number of DPs in 2006 is 25\% higher than on average. 
The coefficients on $\delta_{2006}$ and $\delta_{2011}$ indicate a positive increase in the number of DPs even without changing expenses. The omission of year dummies would lead to the attribution of this positive effects to the effects of costs change.

\begin{figure}[H]
	\begin{center}
		\includegraphics[scale = 0.5]{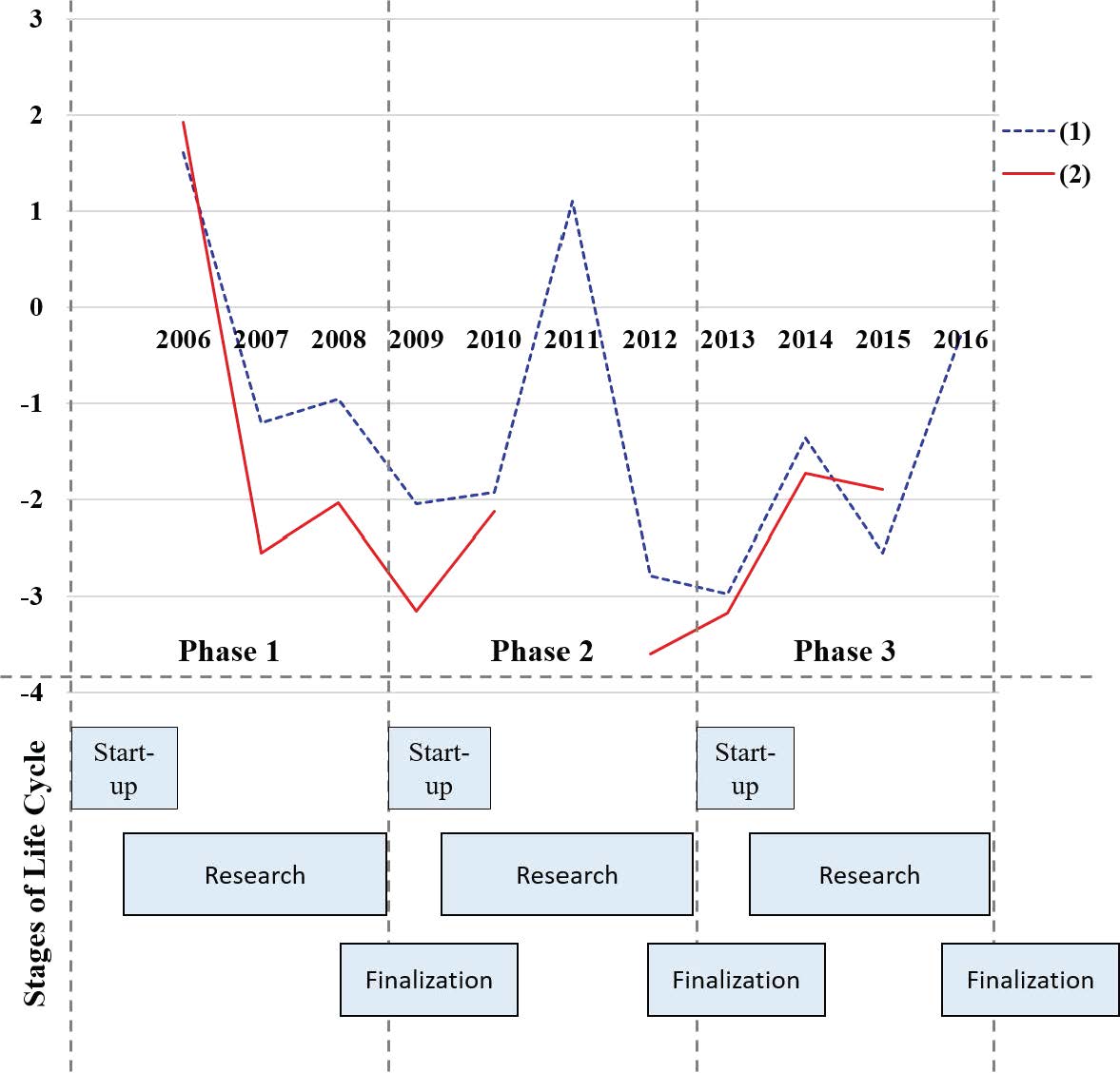}
		\caption{Estimates of coefficients on the year dummy variables for the time fixed effects (within) regression (models (1) and (2)). The lower part of the figure shows the corresponding stage of the research project life cycle.			
		}
		\label{EstLifeCycle1}
	\end{center}
\end{figure}

\begin{figure}[H]
	\begin{center}
		\includegraphics[scale = 0.5]{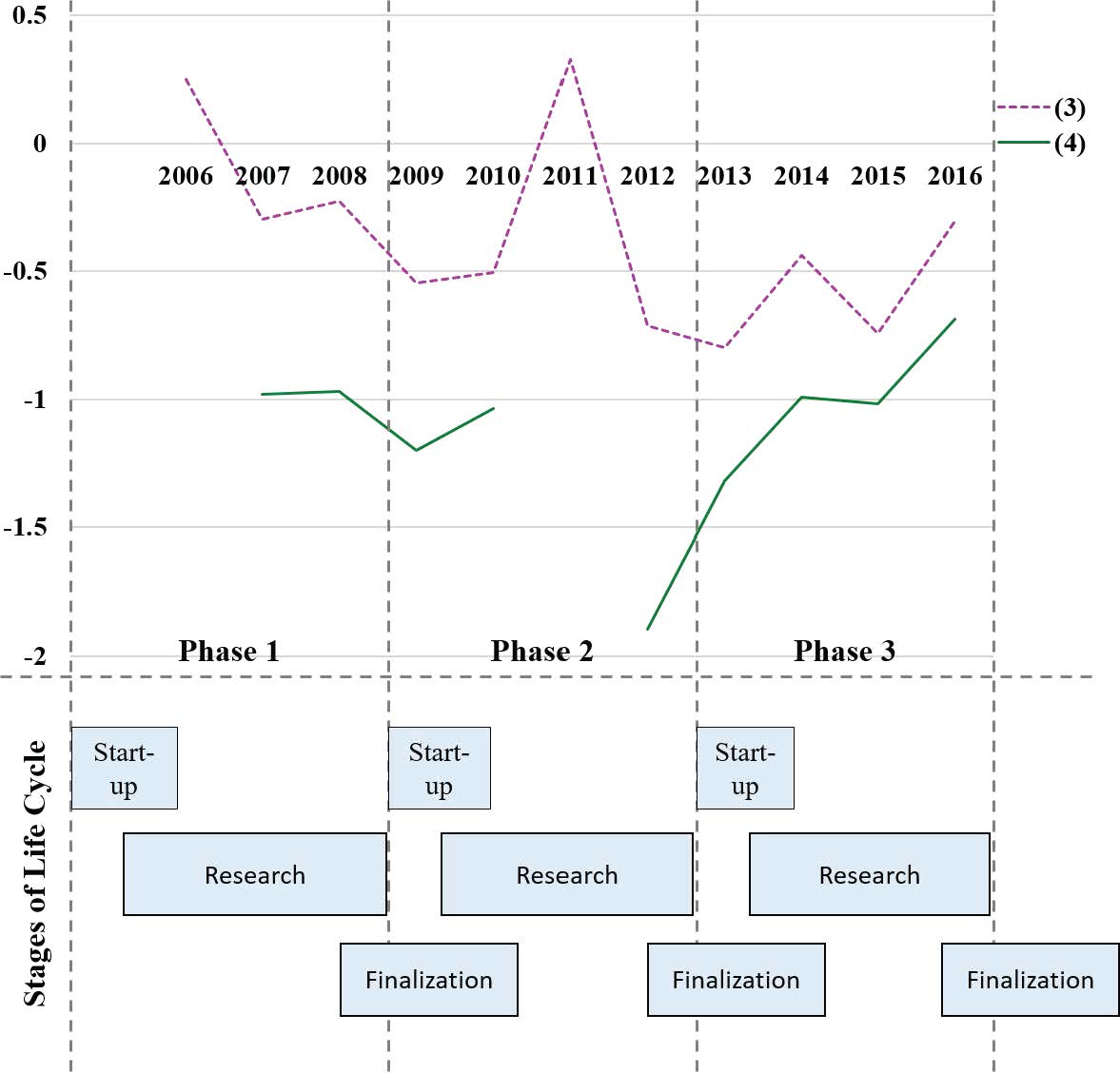}
		\caption{Estimates of coefficients on the year dummy variables for the fixed effects Poisson regression (models (3) and (4)). The lower part of the figure shows the corresponding stage of the research project life cycle.
		}
		\label{EstLifeCycle2}
	\end{center}
\end{figure}

One can see that the year effects have a negative impact on the number of DPs in the majority of years for all models. The project's life cycle could explain this.  
Research projects generally have five main stages: proposal development, funding review, project start-up, performing research and finalization of the project. We map the estimates of coefficients of the models and fit the stages of life cycles in Figures \ref{EstLifeCycle1} and \ref{EstLifeCycle2}. 
Proposal development and funding review take place before 2005 and are not depicted in these Figures.

\vspace{-1cm}

A highly demanding application for a CRC requires extensive preliminary research. The results of this preliminary research are published as DPs in the first year 2005, thus, creating a specific bias towards later research outputs produced during the CRC's life time. The three following increases in the number of DPs take place mainly in the finalization stage caused by the publishing of research results at the final stage of projects. The research outputs of the last phase in 2016 show  part of the positive trend. In fact, 28 DPs were published in 2017, after the CRC was officially finished and financing ended.
Three major declines could be explained by a theoretical and empirical stage of the research in the middle of each project life cycle. 
In summary, the joint depiction of the time fixed effects and the research project's life cycle allows a better understanding of the development of the number of DPs over time.

\section{Conclusions}
\label{sec5}

Our findings show that the performance indicators suitable for the intermediate or final evaluation of a CRC facilitate a better understanding of the dependence structure between research productivity and financial inputs, and provide relevant information for successful decision and policy making.

As a result of semantic analysis of the text from proposals for the CRC submitted to the DFG and the abstracts from published DPs, we find out that two word clouds standing for goals and results use 50\% of the same words. Aiming to visualize a further career path of young researchers that received their Ph.D. within the CRC, we use  mosaic plot with dimensions gender, location and area of work. We show that almost 37\% are females  and 70\% of young researchers found a job in academia. 

We describe the interdisciplinary structure of research results with the help of the network analysis. We show that such fields as mathematical and quantitative methods, financial economics, macroeconomics and monetary economics and microeconomics are the most reflected in the published DPs. These fields correspond to the primary research areas of the CRC. Moreover, the most of research output takes place in the areas that have more PIs with corresponding expertise. Additionally, the sub-projects with more research staff are expected to produce more DPs. The network visualization  provides also the evidence that one of the main goals of the interdisciplinary research center -- interdisciplinarity -- is achieved. 

Using time fixed effects panel data model and fixed effects Poisson model, we show that increasing staff costs by 100\% raises the number of DPs of a sub-project by 1.62 or 43\% according to the estimates of FE and FEP models correspondingly. Travel costs have diminishing effect on the number of DPs according to our estimation results. We analyse the change in productivity of the CRC over time for reasons not captured by the other independent variables using the dummy variables for years. We depict the estimates of coefficients for years and show the possible association between the trend and the stages of a project's life cycle. For instance, the major declines in the number of DPs take place during the stage of theoretical and empirical research, whereas the finalization stage may correspond to the growth in the number of published DPs.


\bibliographystyle{nonumber}

%% Non-BibTeX users please use
%\begin{thebibliography}{}
%	%
%	% and use \bibitem to create references. Consult the Instructions
%	% for authors for reference list style.
%	%
%	\bibitem{RefJ}
%	% Format for Journal Reference
%	Author, Article title, Journal, Volume, page numbers (year)
%	% Format for books
%	\bibitem{RefB}
%	Author, Book title, page numbers. Publisher, place (year)
%	% etc
%\end{thebibliography}

\end{document}